\documentclass[12pt]{article}
\usepackage{epsfig}

\begin{document}

\begin{center}

\vspace*{1.0cm}

{\Large \bf{First search for double $\beta$ decay of dysprosium}}

\vskip 1.0cm

{\bf P.~Belli$^{a}$, R.~Bernabei$^{a,b,}$\footnote{Corresponding
author. {\it E-mail address:} rita.bernabei@roma2.infn.it
(R.~Bernabei).}, F.~Cappella$^{c,d}$, R.~Cerulli$^{e}$,
F.A.~Danevich$^{f}$, S.~d'Angelo$^{a,b}$, M.L.~Di Vacri$^{e}$,
A.~Incicchitti$^{c,d}$, M.~Laubenstein$^{e}$, S.S.~Nagorny$^{f}$,
S.~Nisi$^{e}$, A.V.~Tolmachev$^{g}$, V.I.~Tretyak$^{f}$,
R.P.~Yavetskiy$^{g}$}

\vskip 0.3cm

$^{a}${\it INFN sezione Roma ``Tor Vergata'', I-00133 Rome, Italy}

$^{b}${\it Dipartimento di Fisica, Universit$\grave{a}$ di Roma
``Tor Vergata'', I-00133 Rome, Italy}

$^{c}${\it INFN sezione Roma ``La Sapienza'', I-00185 Rome, Italy}

$^{d}${\it Dipartimento di Fisica, Universit$\grave{a}$ di Roma
``La Sapienza'', I-00185 Rome, Italy}

$^{e}${\it INFN, Laboratori Nazionali del Gran Sasso, 67010
Assergi (AQ), Italy}

$^{f}${\it Institute for Nuclear Research, MSP 03680 Kyiv,
Ukraine}

$^{g}${\it Institute for Single Crystals, 61001 Kharkiv, Ukraine}

\end{center}

\vskip 0.5cm

\begin{abstract}

A search for double $\beta$ decay of dysprosium was realized for
the first time with the help of an ultra low-background HP Ge
$\gamma$ detector. After 2512 h of data taking with a 322 g sample
of dysprosium oxide limits on double beta processes in $^{156}$Dy
and $^{158}$Dy have been established on the level of $T_{1/2}\geq
10^{14}-10^{16}$ yr. Possible resonant double electron captures in
$^{156}$Dy and $^{158}$Dy were restricted on a similar level. As
a by-product of the experiment we have measured the radioactive
contamination of the Dy$_2$O$_3$ sample and set limits on the $\alpha$
decay of dysprosium isotopes to the excited levels of daughter
nuclei as $T_{1/2}\geq 10^{15} - 10^{17}$ yr.

\end{abstract}

\vskip 0.4cm


\vskip 0.4cm


\section{INTRODUCTION}

The double beta ($2\beta$) decay experiments are considered to-date as
the best way to determine an absolute scale of the neutrino mass and to
establish the neutrino mass hierarchy, to clarify the nature of the
neutrino (Majorana or Dirac particle), to look for existence of
right-handed admixtures in the weak interaction and of hypothetical
Nambu-Goldstone bosons (Majorons), and to test some other effects
beyond the Standard Model \cite{DBD-rev}. The developments in the
new experimental techniques during the last two decades lead to an
impressive improvement of sensitivity to the neutrinoless ($0\nu$)
mode of $2\beta^-$ decay up to $10^{23}-10^{25}$ yr
\cite{DBD-tab}. Allowed in the Standard Model the two neutrino
($2\nu$) double beta decay was detected for 10 nuclides with the
half-lives in the range of $10^{18}-10^{24}$ yr
\cite{DBD-tab,Bar10}.

The sensitivity of the experiments to search for the double electron
capture ($2\varepsilon$), the electron capture with emission of
positron ($\varepsilon\beta^{+}$), and the double positron
($2\beta^{+}$) decay is substantially lower: the best counting
experiments give only limits on the level of $10^{18}-10^{21}$ yr
\cite{DBD-tab,Gav06,Bel09,Ruk10}\footnote{An indication for
$2\beta ^{+}$ decay processes in $^{130}$Ba and $^{132}$Ba was
obtained in geochemical measurements \cite{Mesh01}; however, this
result has to be confirmed in a direct counting experiment.}.
There is a strong motivation to develop experimental technique to
search for these processes: study of neutrinoless $2\varepsilon$
and $\varepsilon\beta^+$ decays could clarify the contribution of
the right-handed admixtures in weak interactions \cite{Hir94}.
Dysprosium contains two potentially $2\beta$ active isotopes:
$^{156}$Dy with one of the largest releases $Q_{2\beta}=(2012\pm6)$
keV (therefore both $2\varepsilon$ and $\varepsilon\beta^+$
channels of decay are possible), and $^{158}$Dy
($Q_{2\beta}=(284.6\pm2.5)$ keV, only double electron capture is
energetically allowed) \cite{Aud03}. 
The decay schemes of the
triplets $^{156}$Dy$~-^{156}$Tb$~-^{156}$Gd and
$^{158}$Dy$~-^{158}$Tb$~-^{158}$Gd are presented in Fig. 1 and 2,
respectively.
\begin{figure}[!htb]
\begin{center}
 \mbox{\epsfig{figure=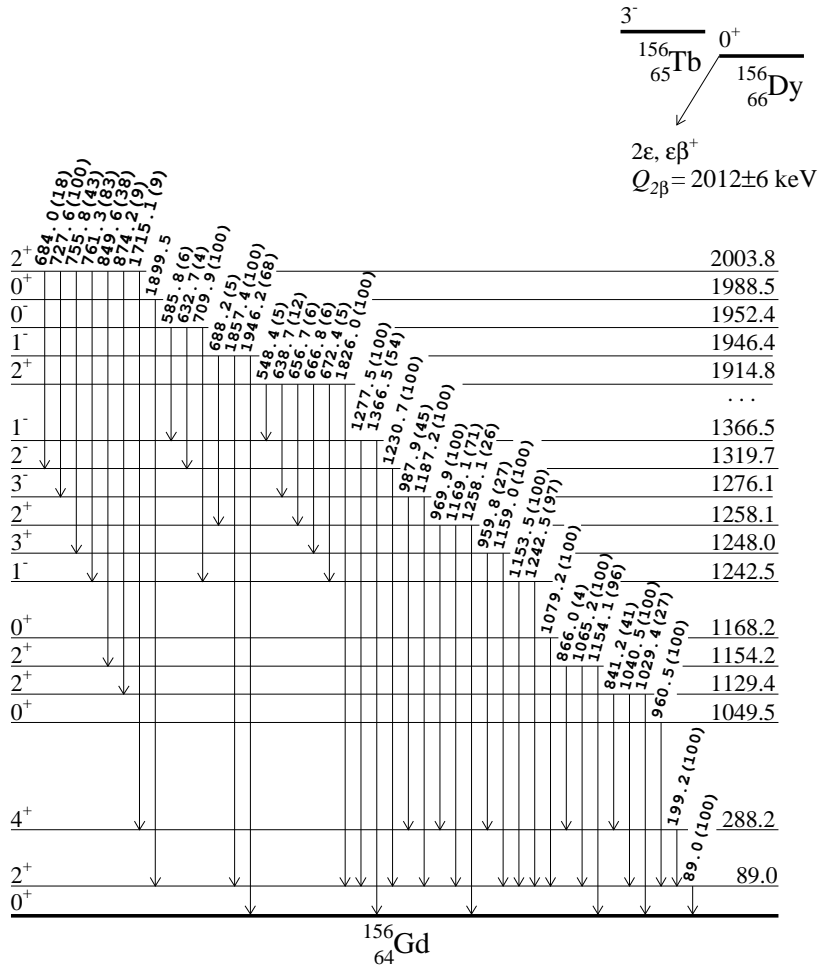,height=13.0cm}}
 \caption{Decay scheme of $^{156}$Dy \cite{Rei03}. Energies of excited levels and emitted
$\gamma$ quanta are in keV (relative intensities of $\gamma$
quanta, roughed to percent, are given in parentheses). Excited
states related with possible population and subsequent
de-excitation of $0^+$ levels at $E_{exc}=1851.2$ and 1715.2 keV,
and $2^+$ levels at $E_{exc}=1827.8$ and 1771.1 keV are not shown.
Deexcitation branches with low probabilities ($\le1\%$) also are
not shown.}
\end{center}
\end{figure}

\begin{figure}[!htb]
\begin{center}
 \mbox{\epsfig{figure=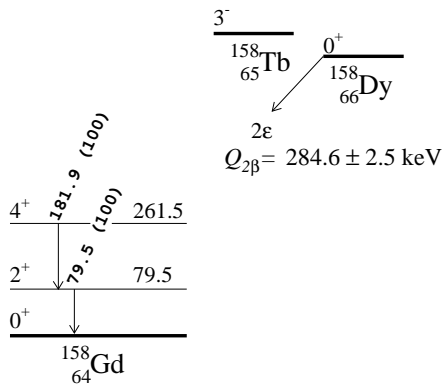,height=5.0cm}}
\caption{Decay scheme of $^{158}$Dy \cite{Hel04}. Energy of
excited levels and $\gamma$ quanta are in keV.}
\end{center}
\vspace{-0.4cm}
\end{figure}

It should be mentioned the possibility of a resonant enhancement of
the neutrinoless double electron capture in $^{156}$Dy and
$^{158}$Dy due to energy degeneracy. The resonant double electron
capture was discussed in Refs. \cite{Win55,Vol82,Ber83,Suj04},
where an enhancement of the decay rate by some orders of
magnitude was predicted for the case of coincidence between the
released energy and the energy of an excited state. According to
\cite{Suj04}, high $Z$ atoms are strongly favored to search for
resonant $2\varepsilon$ decay. Dysprosium has one of the highest
$Z$ among nuclides for which resonant processes could occur.

Resonant captures are possible on a few excited levels of
$^{156}$Gd and one level of $^{158}$Gd. The properties of the
excited levels are listed in Table~\ref{tb1}. 
\begin{table}[htb]
\caption{Characteristics of possible resonant neutrinoless double
electron capture in $^{156}$Dy and $^{158}$Dy. Difference
$Q_{2\beta}-E_{exc}$ (where $Q_{2\beta}$ is the double beta decay
energy, $E_{exc}$ is the energy of an excited level) is denoted as
$\delta$.}
\begin{center}
\begin{tabular}{|l|l|l|l|}
 \hline
  Transition                        & Decay channel & Energy (keV) and parity   & $\delta$ (keV) \\
   ~                                & ~             & of excited level          & ~              \\
  \hline
  $^{156}$Dy$\rightarrow$$^{156}$Gd & $2K$          & 1914.8, 2$^+$             & $-3\pm 6$     \\
  ~                                 & $KL_1$        & 1946.4, 1$^-$             & $~~7\pm 6$ \\
  ~                                 & $KL_1$        & 1952.4, 0$^-$             & $~~1\pm 6$ \\
  ~                                 & $2L_1$        & 1988.5, 0$^+$             & $~~7\pm 6$ \\
  ~                                 & $2L_3$        & 2003.8, 2$^+$             & $-6\pm 6$ \\
  \hline
  $^{158}$Dy$\rightarrow$$^{158}$Gd & $2L_1$        & 261.5, 4$^+$              & $~~6.3\pm 2.5$     \\
 \hline
\end{tabular}
\end{center}
\label{tb1}
\end{table}
Because transitions with
difference in spin more than 2 are strongly suppressed, we
consider in this study only the levels of $^{156}$Gd with spin
$\leq2$. However, we left in the list the level of $^{158}$Gd with
the spin 4$^+$ to which the resonant capture is possible.

Unfortunately, the isotopic abundances of both potentially double
beta active dysprosium isotopes are rather low: concentrations of
$^{156}$Dy and $^{158}$Dy in the natural dysprosium are 0.056(3)\%
and 0.095(3)\%, respectively \cite{Boh05}.

To our knowledge there were no attempts yet to search for double
$\beta$ decays of $^{156}$Dy and $^{158}$Dy. The aim of the
present work was the search for $2\beta$ processes in the dysprosium
isotopes with the help of ultra-low background high purity (HP) Ge 
$\gamma$
spectrometry. As a by-product of the experiment we have estimated
the radioactive contamination of the dysprosium oxide sample and set
limits on the $\alpha$ decay of the dysprosium isotopes to excited levels
of the daughter nuclei.

\section{MEASUREMENTS, RESULTS AND DISCUSSION}

\subsection{Contamination of Dy$_2$O$_3$ measured by mass-spectrometry}

A sample of high pure dysprosium oxide (Dy$_2$O$_3$ of 99.98\%
purity grade) was provided by the Institute for Single Crystals
(Kharkiv, Ukraine). The contamination of the material was measured
with the help of High Resolution Inductively Coupled Plasma-Mass
Spectrometric analysis (HR-ICP-MS, Thermo Fisher Scientific
ELEMENT2). Potassium was measured in High Resolution mode.
The calibration of the apparatus was performed by adding in one of the
sample solutions of a known amount of a standard solution
containing this element. All the other elements were determined in
Low Resolution High Sensitivity mode. 
\begin{table}[htb]
\caption{Contamination of Dy$_2$O$_3$ analyzed by ICP-MS analysis.
The calculated activities of the most important radioactive isotopes
presented in the radioactive elements are given.}
\begin{center}
\begin{tabular}{|l|l|l|}
 \hline
  Element           & Concentrations    &  Activity of   \\
  ~                 &  of element       &  radioactive isotopes \\
  ~                 & (ppm)             & (mBq/kg)      \\
  ~                 & ~                 & ~ \\
  \hline
  K                 & $\leq0.5$         & $\leq15$ ($^{40}$K) \\
  Ca                & 50                & ~ \\
  Co                & $\leq 0.05$         & ~ \\
  Cu                & $\leq 5$          & ~ \\
  Sr                & $\leq 0.02$       & ~ \\
  Y                 & 6                 & ~ \\
  La                & 0.25              & 0.21 ($^{138}$La) \\
  Sm                & 17                & 2200 ($^{147}$Sm) \\
  Gd                & 5                 & 0.008 ($^{152}$Gd) \\
  Tb                & $\leq 5$          & ~ \\
  Ho                & 30                & ~ \\
  Er                & 60                & ~ \\
  Yb                & 3                 & ~ \\
  Lu                & 0.5               & 26  ($^{176}$Lu) \\
  Pb                & 0.16              & ~ \\
  Th                & 0.036             & 147 ($^{232}$Th) \\
  U                 & 0.004             & 2.3 ($^{235}$U), 49 ($^{238}$U)  
\\
 \hline
\end{tabular}
\end{center}
\label{tb2}
\end{table}
A semiquantitative
analysis was performed, that is a single standard solution
containing some elements at a known concentration level (10 ppb of
Li, Ce, Tl) was used for calibration and quantification. In this
case, the uncertainty can be estimated as about 30\% of the given
concentration value. The results of the HR-ICP-MS analysis are
presented in Table \ref{tb2}. We have also calculated the activity of
the radionuclides present in the measured elements.

\subsection{Low background measurements}

A Dy$_2$O$_3$ sample with the mass of 322 g in a polyethylene bag
was placed directly on the end cap of the ultra-low background HP
Ge detector (GeBer, 244 cm$^3$) installed deep underground
($\simeq$3600 meters of water equivalent) at the Gran Sasso
National Laboratories of the INFN (Italy). The detector was
located inside a passive shield made of low radioactivity
lead ($\approx 20$ cm), copper ($\approx 10$ cm) and borated
polyethylene ($\approx 10$ cm). To remove radon, the set-up was
continuously flushed by highly pure nitrogen. The energy resolution
of the spectrometer is 2 keV at 1332 keV $\gamma$ line of
$^{60}$Co. The Dy$_2$O$_3$ sample was measured over 2512 h, and
the background of the detector was collected during 6110 h.

The energy spectrum accumulated with the Dy$_2$O$_3$ sample by the
HP Ge detector is shown in Fig. 3 together with the background
data. 
\begin{figure}[htb]
\begin{center}
 \mbox{\epsfig{figure=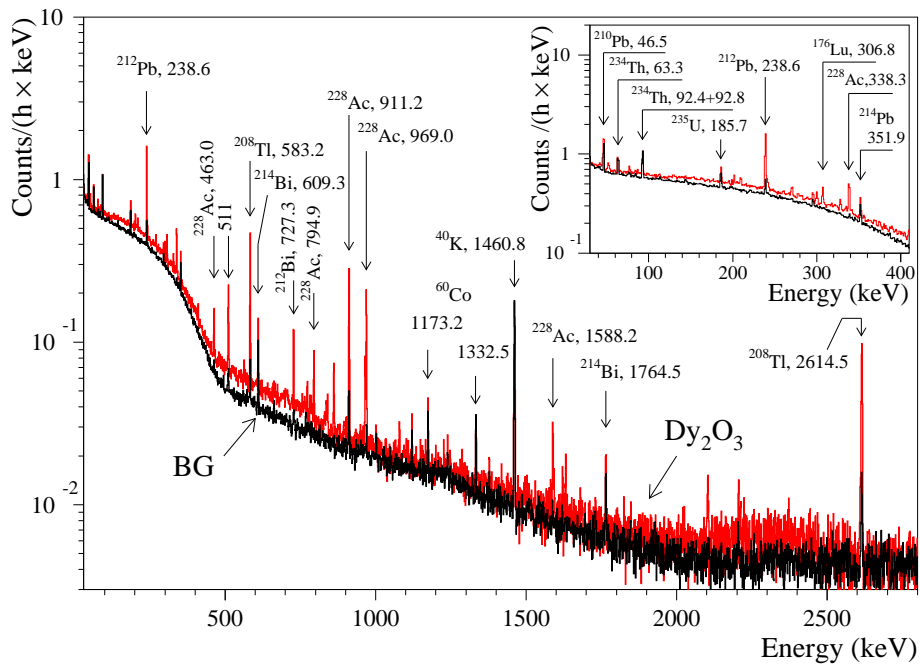,height=9.0cm}}
 \caption{(Color online) Energy spectra accumulated with dysprosium
sample over 2512 h (Dy$_2$O$_3$) and without sample over 6110 h
(BG) by ultra-low background HP Ge $\gamma$ spectrometer. (Inset)
Low energy part of the spectra. Energy of $\gamma$ lines are in
keV.}
\end{center}
\end{figure}
The spectra are normalized on the times of measurements.
The peaks in the spectra can be ascribed to $\gamma$ quanta of
the radionuclides $^{40}$K, $^{60}$Co, and
daughters of U/Th. We have also observed $\gamma$ lines of
$^{176}$Lu (201.8 and 306.8 keV) in the data accumulated with the
dysprosium sample. To estimate radioactive contamination of the
Dy$_2$O$_3$ sample, the detection efficiencies of the background
components were simulated with the help of the EGS4 package
\cite{EGS4}. The initial kinematics of the particles emitted in
the decay of nuclei was given by an event generator DECAY0
\cite{DECAY0}. Radioactive contamination of the dysprosium oxide
is presented in 
Table \ref{tb3}. 
\nopagebreak
\begin{table}[htb]
\caption{Radioactive contamination of Dy$_2$O$_3$ sample measured
by HP Ge $\gamma$ spectrometry.}
\begin{center}
\begin{tabular}{|l|l|l|}

 \hline
  Chain     & Nuclide       & Activity (mBq/kg) \\
 \hline
 ~          & $^{40}$K      & $\leq 10$  \\
 ~          & $^{60}$Co     & $\leq 1$  \\
 ~          & $^{137}$Cs    & $\leq 2$ \\
 ~          & $^{138}$La    & $\leq 3$ \\
 ~          & $^{176}$Lu    & $9\pm2$ \\
 \hline
 $^{232}$Th & $^{228}$Ra    & $179\pm40$ \\
 ~          & $^{228}$Th    & $157\pm15$ \\
 \hline
 $^{235}$U  & $^{235}$U     & $7\pm3$ \\
 ~          & $^{231}$Pa    & $\leq 80$ \\
 ~          & $^{227}$Ac    & $\leq 190$ \\
 \hline
 $^{238}$U  & $^{234m}$Pa   & $\leq 115$ \\
  ~         & $^{226}$Ra    & $5\pm3$  \\
 ~          & $^{210}$Pb    & $\leq 9400$  \\
  \hline

\hline
\end{tabular}
\end{center}
\label{tb3}
\end{table}
The data agree with the results of the
ICP-MS analysis taking into account accuracy of the measurements.

The observed contamination of the sample by U/Th is quite
predictable taking into account that the principal ores for
production of dysprosium are monazite and bastnaesite. Monazite
contains considerable (typically a few \%) amount of thorium and
uranium. The presence of radioactive $^{176}$Lu can also be expected
due to the very similar chemical properties of lutetium and
dysprosium, which lead to certain difficulties in chemical
separation of the elements.

\subsection{Search for $2\varepsilon$ and
$\varepsilon\beta^+$ decay of $^{156}$Dy}

There are no clear peculiarities in the energy spectra accumulated
with the dysprosium sample, which can be interpreted as double
beta decay of $^{156}$Dy or $^{158}$Dy. Therefore only lower
half-life limits can be set according to the formula:

\begin{center}
$$\lim T_{1/2} = N \cdot \eta \cdot t \cdot \ln 2 / \lim S,$$
\end{center}
where $N$ is the number of potentially $2\beta$ unstable nuclei,
$\eta$ is the detection efficiency, $t$ is the measuring time, and
$\lim S$ is the number of events of the effect searched for which
can be excluded at a given confidence level (C.L.).

To estimate values of $\lim S$ for $2\beta$ processes in
$^{156}$Dy and $^{158}$Dy, the energy spectrum accumulated with
the Dy$_2$O$_3$ sample was fitted in different energy regions
where $\gamma$ peaks related with the decay processes are
expected. The efficiencies of the HP Ge detector for the double
$\beta$ processes searched for were simulated with the help of the
EGS4 code \cite{EGS4} and the DECAY0 event generator
\cite{DECAY0}.

One positron can be emitted in the $\varepsilon\beta^+$ decay of
$^{156}$Dy with the maximal energy of ($990\pm6$) keV. The annihilation
of the positron will give two 511 keV $\gamma$'s leading to extra
counting rate in the annihilation peak. The energy spectra
accumulated with and without the sample in the energy interval
($400-700$) keV are presented in Fig. 4.
\begin{figure}[htb]
\begin{center}
 \mbox{\epsfig{figure=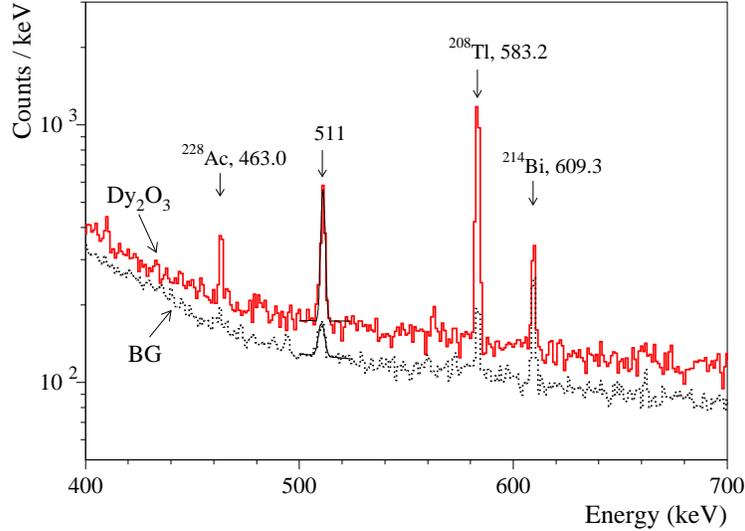,height=7.5cm}}
\caption{(Color online) Part of energy spectra accumulated with
Dy$_2$O$_3$ sample over 2512 h by ultra-low background HP Ge
$\gamma$ spectrometer (Dy$_2$O$_3$). The energy spectrum
accumulated without sample over 6110 h normalized on the time of
measurements with the Dy$_2$O$_3$ sample (BG). Fits of 511
annihilation $\gamma$ peaks are shown by solid lines.}
\end{center}
\end{figure}
The measured area of the
annihilation peak is ($862\pm44$) counts. The area of the peak in
the background spectrum is ($149\pm23$) counts (normalized on the
time of measurement with the Dy$_2$O$_3$ sample). The excess of
events in the data accumulated with the dysprosium sample can be
explained by radioactive contamination of the material. Indeed
decays of $^{226}$Ra and $^{228}$Th daughters give in total
($757\pm72$) counts in the annihilation peak. The difference between
the observed and the expected number of counts ($-44\pm87$) gives
no evidence for the effect of the $\varepsilon\beta^+$ decay of
$^{156}$Dy. In accordance with the Feldman-Cousins procedure
\cite{Fel98}, we obtain 102 counts which can be excluded at 90\%
C.L. (here and below all the half-life limits and values of $\lim
S$ are given at 90\% C.L.). Taking into account the number of
$^{156}$Dy nuclei in the sample ($N=2.91\times10^{20}$) and the
detection efficiency (3.4\%), we have obtained the following limit
on the half-life of $^{156}$Dy relatively to the sum of $2\nu$ and
$0\nu$ modes of $\varepsilon\beta^+$ decay:

\begin{center}
 $T_{1/2}^{(2\nu+0\nu)\varepsilon \beta^+}(^{156}$Dy, g.s.$~\rightarrow~$g.s.$)\geq 1.9\times10^{16}$ yr.
\end{center}

The $\varepsilon\beta^+$ decay of $^{156}$Dy is also allowed to
the first excited level of $^{156}$Gd with the energy of 89.0 keV.
To estimate an upper limit on the transition, the energy spectrum
was fitted in the energy interval ($83-109$) keV by the model
constructed from a Gaussian function at the energy of 89 keV with
the energy resolution FWHM$~=1.6$ keV (the $\gamma$ peak searched
for), a linear function describing the background, and a second
Gaussian to take into account the $\gamma$ peaks with the energies
92.4 and 92.8 keV ($^{234}$Th from $^{238}$U chain). A fit by the
chisquare method ($\chi^{2}/$n.d.f.$~= 18.9/19=0.99$, where n.d.f.
is number of degrees of freedom) gives the area of the peak
searched for: $S=(86\pm88)$ counts, which gives no evidence for the
effect. We took 230 counts which can be excluded at 90\% C.L.
Taking into account the simulated efficiency to 89 keV $\gamma$
quanta (0.05\%) we have obtained the limit on
$2\nu\varepsilon\beta^+$ decay of $^{156}$Dy to the first $2^+$
89.0 keV excited level of $^{156}$Gd:
$T_{1/2}^{2\nu\varepsilon\beta^{+}}($g.s.$~\rightarrow~89.0~$keV$)\geq~1.3\times10^{14}$
yr. However, a much stronger restriction can be set by considering
the annihilation peak. Indeed, the detection efficiency for the 511
keV $\gamma$ for the $2\nu\varepsilon\beta^{+}$ (and
$0\nu\varepsilon\beta^{+}$) decay to the 89.0 keV excited level is
3.4\% which leads to the same limits as for the
$\varepsilon\beta^{+}$ decay to the ground state:

\begin{center}
 $T_{1/2}^{(2\nu+0\nu)\varepsilon \beta^+}(^{156}$Dy, g.s.$~\rightarrow~$89.0 keV$)\geq 1.9\times10^{16}$ yr.
\end{center}

In a case of the $2\nu2K$ capture in $^{156}$Dy a cascade of X
rays and Auger electrons with the individual energies up to 50 keV
is expected. The most intensive X ray lines are 42.3 keV (26.6\%),
43.0 keV (47.5\%), 48.6 keV (4.8\%), 48.7 keV (9.3\%) and 50.0 keV
(3.1\%) \cite{ToI98}. To derive a limit on the two neutrino double
electron capture from $K$ shells, the energy spectrum was fitted
in the energy interval ($35-60$) keV by the model consisting of five
Gaussians (to describe the expected X ray peaks) plus two
Gaussians (46.5 keV $\gamma$ line of $^{210}$Pb and 53.2 keV
$\gamma$ line of $^{234}$U), and a polynomial function of the
first degree (background; see Fig. 5). The fit gives the total area of the
$2\nu2K$ effect as ($310\pm102$) counts. It is worth noting that the background in
this energy region is quite complex to be safely estimated; therefore,
conservatively we give here
only a limit on the number of events: $\lim S=477$ counts. Taking
into account the detection efficiency of the effect (0.5\%), one
can calculate the following half-life limit:

 \begin{center}
 $T_{1/2}^{2\nu2K}(^{156}$Dy, g.s.$~\rightarrow~$g.s.$)\geq 6.1\times10^{14}$ yr.
 \end{center}

To estimate a limit on the $2\nu2\varepsilon$ decay of $^{156}$Dy
to the excited levels of $^{156}$Gd, the energy spectrum
accumulated with the dysprosium sample was fitted in energy
intervals where peaks from the de-excitation $\gamma$ rays are
expected. 
\begin{figure}[!htb]
\vspace{-0.3cm}
\begin{center}
 \mbox{\epsfig{figure=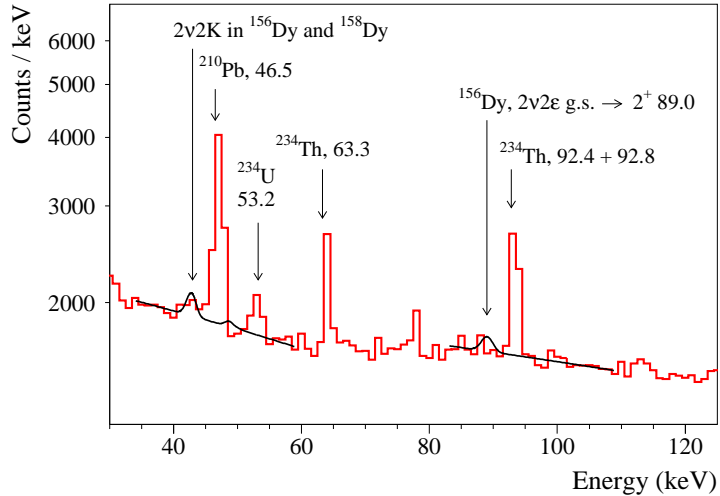,height=7.0cm}}
\caption{(Color online) Low energy part of energy spectrum
accumulated with Dy$_2$O$_3$ sample over 2512 h. An excluded
effect of $2\nu2K$ process in $^{156}$Dy ($^{158}$Dy) with the
half-life $T_{1/2}=6.1\times10^{14}$ yr
($T_{1/2}=1.0\times10^{15}$ yr), and $2\nu2\varepsilon$ transition
of $^{156}$Dy to the first excited level of $^{156}$Gd (89.0 keV,
$T_{1/2}=1.8\times10^{14}$ yr) are shown by solid lines.}
\end{center}
\vspace{-0.3cm}
\end{figure}
For instance, we have obtained the following limit on
the $2\nu2\varepsilon$ decay of $^{156}$Dy to the first excited
89.0 keV level of $^{156}$Gd:

 \begin{center}
 $T_{1/2}^{2\nu2\varepsilon}(^{156}$Dy, g.s.$~\rightarrow~89.0~$keV$)\geq 1.8\times10^{14}$ yr.
 \end{center}

A peak expected for the $2\nu2\varepsilon$ transition of
$^{156}$Dy to the excited 2$^+$ 89.0 keV level of $^{156}$Gd is
shown in Fig. 5.

The limits obtained for the $2\nu2\varepsilon$ decay of $^{156}$Dy
to the excited levels of $^{156}$Gd are presented in Table \ref{tb4} where
the energies of $\gamma$ quanta used in the analysis, detection
efficiencies and values of $\lim S$ are given too.

In case of $0\nu2\varepsilon$ decay to the ground state of
$^{156}$Gd, we suppose here that only one bremsstrah\-lung
$\gamma$ quantum is emitted to carry out the transition energy (in
addition to X rays and Auger electrons from de-excitation of
atomic shell). The energy of the $\gamma$ quantum is expected to
be equal to $E_\gamma=Q_{2\beta}-E_{b1}-E_{b2}$, where $E_{b1}$
and $E_{b2}$ are the binding energies of the first and of the
second captured electron of the atomic shell. The binding
energies on the $K$ and $L_1, L_2, L_3$ shells in gadolinium atom
are equal to $E_K=50.2$, $E_{L_1}=8.4$, $E_{L_2}=7.9$ and
$E_{L_3}=7.2$ keV \cite{ToI98}, respectively. Therefore, the
expected energies of the $\gamma$ quanta for the
$0\nu2\varepsilon$ capture in $^{156}$Dy to the ground state of
$^{156}$Gd are: i)$E_\gamma=(1912\pm6)$ keV for $0\nu 2K$; ii)
$E_\gamma=(1954\pm7$) keV for $0\nu KL$; iii)
$E_\gamma=(1996\pm7)$ keV for $0\nu 2L$.

There are no clear peaks in the energy intervals ($1906-1918$),
($1947-1959$), and ($1989-2004$) keV expected in the g.s.
$\rightarrow$ g.s. $0\nu2\varepsilon$ decay of $^{156}$Dy (see
Fig. 6).
\begin{figure}[htb]
\begin{center}
 \mbox{\epsfig{figure=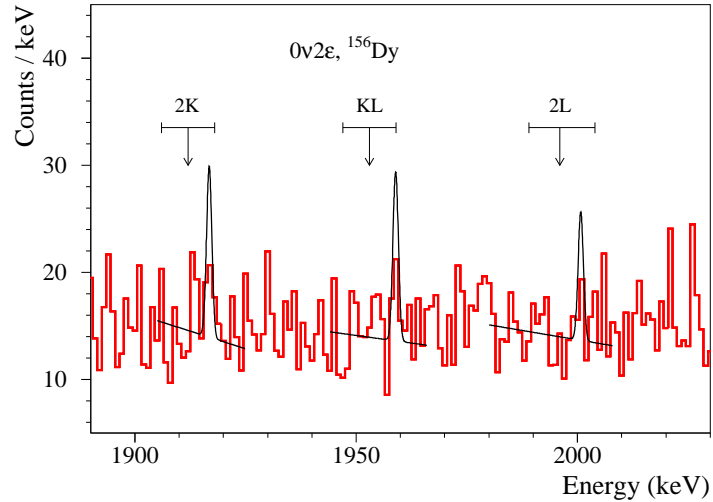,height=7.0cm}}
\caption{(Color online) Part of the energy spectra where peaks
from $0\nu2\varepsilon$ processes ($2K$, $KL$ and $2L$) in
$^{156}$Dy to the ground state of $^{156}$Gd are expected. The
excluded peaks for the processes are shown.}
\end{center}
\end{figure}
\vspace{-0.3cm}
We have estimated values of $\lim S$ by a fit of the data
in the energy intervals by the simple model: a Gaussian function
(to describe the peaks searched for) plus polynomial function
(continuous background). Taking into account the calculated
efficiencies to detect $\gamma$ quanta with energies
$\approx1.9-2.0$ MeV ($\approx0.7\%$), we set the following limits
on the processes searched for:

 \begin{center}
 $T_{1/2}^{0\nu 2K}(^{156}$Dy, g.s.$~\rightarrow~$g.s.$)\geq~1.7\times10^{16}$ yr,

 $T_{1/2}^{0\nu KL}(^{156}$Dy, g.s.$~\rightarrow~$g.s.$)\geq~1.7\times10^{16}$ yr,

 $T_{1/2}^{0\nu 2L}(^{156}$Dy, g.s.$~\rightarrow~$g.s.$)\geq~2.2\times10^{16}$ yr.
 \end{center}

The high $Q_{2\beta}$ energy of $^{156}$Dy allows also population
of several excited levels of $^{156}$Gd with energies in the range
of ($89-2004$) keV. To estimate limits on these processes, the
energy spectrum accumulated with the dysprosium sample was
analyzed in different energy intervals. A summary of the obtained
results is given in Table \ref{tb4}.

\subsection{$2\varepsilon$ decay of $^{158}$Dy}

An estimation of the half-life of $^{158}$Dy relatively to the two
neutrino double $K$ electron capture was obtained by using the
$\lim S$ and the detection efficiency for the $2\nu2K$ decay of
$^{156}$Dy. Taking into account the number of $^{158}$Dy nuclei in
the sample ($N=4.94\times10^{20}$) we have set the following limit
on the $2\nu2K$ decay of $^{158}$Dy:

\begin{center}
 $T_{1/2}^{2\nu2K}(^{158}$Dy, g.s.$~\rightarrow~$g.s.$)\geq 1.0\times 10^{15}$ yr.
\end{center}

The simulated response of the detector to the $0\nu2\varepsilon$
decay of $^{158}$Dy to the ground state of $^{158}$Gd differs from
the two neutrino process. We have set the following limit by
analyzing the data in the vicinity of the 184.2 keV $\gamma$ peak expected
for the $0\nu2K$ decay:

\begin{center}
 $T_{1/2}^{0\nu2K}(^{158}$Dy, g.s.$~\rightarrow~$g.s.$)\geq 4.2\times 10^{16}$ yr.
\end{center}

Limits on the $2\nu$ and $0\nu$ modes of double electron capture
to the first excited level 79.5 keV of $^{158}$Gd were obtained by
analysis of the data in the energy region around the energy 79.5
keV (see Table \ref{tb4}).

\subsection{Search for resonant double electron capture in $^{156}$Dy and $^{158}$Dy}

The double electron capture of $^{156}$Dy to the excited levels of
$^{156}$Gd with energies of 1914.8, 1946.4, 1952.4, 1988.5 and
2003.8 keV may occur with higher probability due to a possible
resonant enhancement. The limits on the resonant double electron
capture in $^{156}$Dy from $K$ and $L$ shells were obtained by
analyzing the experimental spectrum in the energy intervals where
the most intensive $\gamma$ peaks from de-excitation of these
levels are expected. For instance, we set the following limit on
the resonant $2K$ capture in $^{156}$Dy to the excited level $2^+$
of $^{156}$Gd with the energy 1914.8 keV (the dysprosium spectrum
was fitted in the vicinity of the $\gamma$ line 1826.0 keV):

\begin{center}
 $T_{1/2}^{(2\nu+0\nu)2\varepsilon}(^{156}$Dy, g.s.$~\rightarrow~1914.8$ 
keV$)\geq 1.1\times 10^{16}$ yr.
\end{center}

To search for a resonant double electron capture in $^{158}$Dy
from two $L_1$ shells to the excited level 261.5 keV of
$^{158}$Gd, the experimental spectrum was fitted in the energy
interval $175-185$ keV. The fit gives an area of a 181.9 keV peak
($-48\pm49$) which corresponds to $\lim S=40$ counts. The detection
efficiency for $\gamma$ quanta with the energy 181.9 keV
irradiated in the $2\varepsilon$ decay is 1.3\%, therefore we can
set the following limit on the process:

\begin{center}
 $T_{1/2}^{(2\nu+0\nu)2\varepsilon}(^{158}$Dy, g.s.$~\rightarrow~$261.5 keV$)\geq 3.2\times 10^{16}$ yr.
\end{center}

\nopagebreak
\begin{table}
\vspace{-0.3cm}
\caption{Half-life limits on 2$\beta$ processes in $^{156}$Dy and
$^{156}$Dy.}
\begin{center}
\begin{tabular}{|l|l|l|l|l|l|l|}
\hline
 Process                   & Decay       & Level of      & $E_\gamma$  & Detection  & $\lim S$  & Experimental \\
 of decay                  & mode        & daughter      & (keV)       & efficiency & ~         & limit (yr) \\
 ~                         & ~           & nucleus       &             &            &  ~        &  ~\\
 ~                         & ~           & (keV)         &             &            &  ~        &  ~\\
 \hline
 $^{156}$Dy$\to$$^{156}$Gd & ~           & ~             & ~           & ~          &  ~        &  ~\\
 $\varepsilon\beta^+$      & $2\nu+0\nu$ & g.s.          & 511         & 3.4\%      & 102       & $\geq1.9\times10^{16}$ \\
 $\varepsilon\beta^+$      & $2\nu+0\nu$ & $2^+$ 89.0    & 511         & 3.4\%      & 102       & $\geq1.9\times10^{16}$ \\
 $2K$                      & $2\nu$      & g.s.          & $42.3-50.0$ & 0.5\%      & 477       & $\geq6.1\times10^{14}$ \\
 $2\varepsilon$            & $2\nu$      & $2^+$ 89.0    & 89.0        & 0.07\%     & 230       & $\geq1.8\times10^{14}$ \\
 $2\varepsilon$            & $2\nu$      & $0^+$ 1049.5  & 960.5       & 1.1\%      & 9         & $\geq7.1\times10^{16}$ \\
 $2\varepsilon$            & $2\nu$      & $2^+$ 1129.4  & 1040.5      & 0.6\%      & 24        & $\geq1.4\times10^{16}$ \\
 $2\varepsilon$            & $2\nu$      & $2^+$ 1154.1  & 1154.1      & 0.5\%      & 62        & $\geq4.7\times10^{15}$ \\
 $2\varepsilon$            & $2\nu$      & $0^+$ 1168.2  & 1079.2      & 1.0\%      & 65        & $\geq8.9\times10^{15}$ \\
 $2\varepsilon$            & $2\nu$      & $0^+$ 1715.2  & 472.7       & 1.1\%      & 21        & $\geq3.0\times10^{16}$ \\
 $2\varepsilon$            & $2\nu$      & $2^+$ 1771.1  & 1682.2      & 0.7\%      & 39        & $\geq1.0\times10^{16}$ \\
 $2\varepsilon$            & $2\nu$      & $2^+$ 1827.8  & 1738.9      & 0.2\%      & 6         & $\geq1.9\times10^{16}$ \\
 $2\varepsilon$            & $2\nu$      & $0^+$ 1851.2  & 697.0       & 0.1\%      & 39        & $\geq1.5\times10^{15}$ \\
 $2K$                      & $0\nu$      & g.s.          & $1906-1918$ & 0.7\%      & 24        & $\geq1.7\times10^{16}$ \\
 $KL$                      & $0\nu$      & g.s.          & $1947-1959$ & 0.7\%      & 24        & $\geq1.7\times10^{16}$ \\
 $2L$                      & $0\nu$      & g.s.          & $1989-2001$ & 0.7\%      & 18        & $\geq2.2\times10^{16}$ \\
 $2\varepsilon$            & $0\nu$      & $2^+$ 89.0    & 89.0        & 0.06\%     & 230       & $\geq1.5\times10^{14}$ \\
 $2\varepsilon$            & $0\nu$      & $0^+$ 1049.5  & 960.5       & 1.0\%      & 9         & $\geq6.4\times10^{16}$ \\
 $2\varepsilon$            & $0\nu$      & $2^+$ 1129.4  & 1040.5      & 0.6\%      & 24        & $\geq1.4\times10^{16}$ \\
 $2\varepsilon$            & $0\nu$      & $2^+$ 1154.1  & 1065.2      & 0.5\%      & 70        & $\geq4.1\times10^{15}$ \\
 $2\varepsilon$            & $0\nu$      & $0^+$ 1168.2  & 1079.2      & 0.9\%      & 65        & $\geq8.0\times10^{15}$ \\
 $2\varepsilon$            & $0\nu$      & $0^+$ 1715.2  & 472.7       & 1.0\%      & 21        & $\geq2.8\times10^{16}$ \\
 $2\varepsilon$            & $0\nu$      & $2^+$ 1771.1  & 1682.2      & 0.6\%      & 39        & $\geq8.9\times10^{15}$ \\
 $2\varepsilon$            & $0\nu$      & $2^+$ 1827.8  & 1738.9      & 0.2\%      & 6         & $\geq1.9\times10^{16}$ \\
 $2\varepsilon$            & $0\nu$      & $0^+$ 1851.2  & 697.0       & 0.1\%      & 39        & $\geq1.5\times10^{15}$ \\
 Resonant $2K$             & $2\nu+0\nu$ & $2^+$ 1914.8  & 1826.0      & 0.5\%      & 27        & $\geq1.1\times10^{16}$ \\
 Resonant $KL_1$           & $2\nu+0\nu$ & $1^-$ 1946.4  & 1857.4      & 0.4\%      & 24        & $\geq9.6\times10^{15}$ \\
 Resonant $KL_1$           & $2\nu+0\nu$ & $0^-$ 1952.4  & 709.9       & 1.2\%      & 27        & $\geq2.6\times10^{16}$ \\
 Resonant $2L_1$           & $2\nu+0\nu$ & $0^+$ 1988.5  & 1899.5      & 0.7\%      & 21        & $\geq1.9\times10^{16}$ \\
 Resonant $2L_3$           & $2\nu$      & $2^+$ 2003.8  & 1715.1      & 0.02\%     & 41        & $\geq2.8\times10^{14}$ \\
 Resonant $2L_3$           & $0\nu$      & $2^+$ 2003.8  & 761.3       & 0.03\%     & 58        & $\geq3.0\times10^{14}$ \\
 ~                         & ~           & ~             & ~           & ~          & ~         & ~                     \\
 $^{158}$Dy$\to$$^{158}$Gd & ~           & ~             & ~           & ~          & ~         & ~                     \\
 $2K$                      & $2\nu$      & g.s.          & $42.3-50.0$ & 0.5\%      & 477       & $\geq1.0\times10^{15}$  \\
 $2K$                      & $0\nu$      & g.s.          & 184.2       & 1.7\%      & 40        & 
$\geq4.2\times10^{16}$  \\
 $2\varepsilon$            & $2\nu$      & $2^+$  79.5   & 79.5        & 0.04\%     & 113       & $\geq3.5\times10^{14}$ \\
 $2\varepsilon$            & $0\nu$      & $2^+$  79.5   & 79.5        & 0.03\%     & 113       & $\geq2.6\times10^{14}$ \\
 Resonant $2L_1$           & $2\nu+0\nu$ & $4^+$ 261.5   & 181.9       & 1.3\%      & 40        & $\geq3.2\times10^{16}$  \\
 \hline
\end{tabular}
\end{center}
\label{tb4}
\end{table}
All the half-life limits on $2\beta$ decay processes in dysprosium
obtained in the present experiment are summarized in Table \ref{tb4}.

It should be stressed that the difference between the decay energy and
the energy of an excited level is one of the most crucial characteristics
for the resonant double electron capture. However, in most of the
cases the accuracy of atomic mass data is still not enough to
point out isotopes with minimal differences $Q_{2\beta}-E_{exc}$.
Therefore searches for resonant transitions in different isotopes
are strongly motivated.

\subsection{Search for $\alpha$ decay of dysprosium isotopes}

Five of the seven natural dysprosium isotopes are potentially
unstable for $\alpha$ decay. Furthermore for all the
isotopes transitions to excited levels of daughter nuclei are
allowed. Characteristics of the possible alpha decays are
presented in Table \ref{tb5}. The search for $\alpha$ decay of the 
dysprosium
isotopes to the excited levels of the daughter gadolinium isotopes
was realized by analysing the low background energy spectrum
measured with the Dy$_2$O$_3$ sample.

We do not observe any peculiarities in the experimental data which
could be interpreted as $\gamma$ peaks from $\alpha$ decays of the
dysprosium isotopes to the lowest excited levels of the daughter
nuclei, and set only limits on these decays. To estimate a limit on the
$\alpha$ decay of $^{156}$Dy to the excited $2^+$ 344.3 keV level
of $^{152}$Gd, the energy spectrum of the Dy$_2$O$_3$ sample was
fitted in the energy interval ($341-350$) keV by a simple model:
Gaussian function (to describe $\gamma$ peak with the energy of
344.3 keV) and polynomial function of the second degree (to
approximate background, see Fig. 7).
\begin{figure}[htb]
\begin{center}
 \mbox{\epsfig{figure=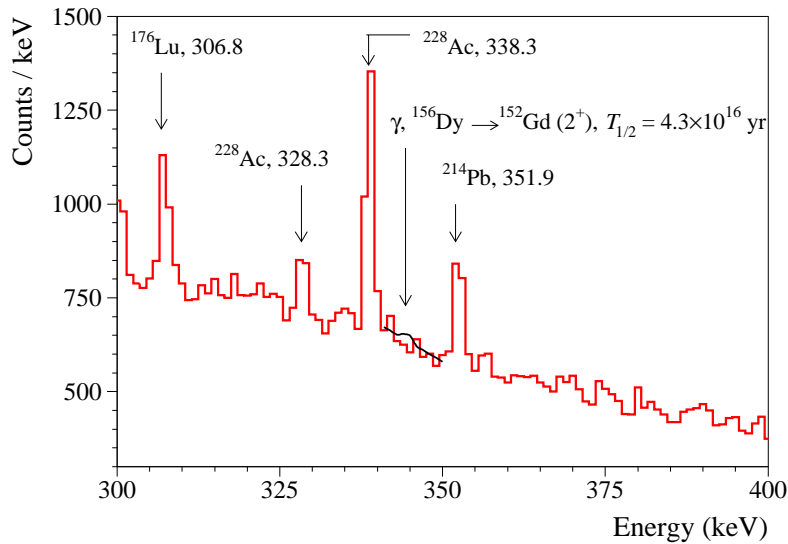,height=7.5cm}}
\caption{(Color online) A part of the energy spectrum of HP Ge
spectrometer accumulated with Dy$_2$O$_3$ sample over 2512 h. An
expected $\gamma$ peak from $\alpha$ decay of $^{156}$Dy to the
first excited level of $^{152}$Gd corresponding to the half-life
$T_{1/2} = 4.3\times10^{16}$ yr (excluded at 90\% C.L.) is shown
by the solid line.}
\end{center}
\end{figure}
A fit by the chisquare
method ($\chi^2$/n.d.f. $= 5.3/5 =1.1$) gives $S=(-48\pm42$) counts
for the area of the peak searched for, which provides no
indication for the effect. According to the Feldman-Cousins
procedure \cite{Fel98} we took 31 counts as $\lim S$. Taking into
account the detection efficiency (2.1\%) and the yield of $\gamma$
quanta from the transitions from the excited level $2^+$ 344.3 keV
(0.962), we have obtained the following limit on the $\alpha$
decay of $^{156}$Dy to the first excited level of $^{152}$Gd:

\begin{center}
 $T_{1/2}^{\alpha}(^{156}$Dy, g.s.$~\rightarrow~$344.3 keV$)\geq 3.8\times 10^{16}$ yr.
\end{center}

Limits on $T_{1/2}$ relatively to $\alpha$ decay to the lowest
excited levels of the daughter nuclei for other dysprosium
isotopes were obtained in the same way. The results are presented
in Table \ref{tb5} where theoretical estimations calculated by using the
approach proposed in \cite{Poe83} are given too. 
\nopagebreak
\begin{table}[!h]
\caption{Characteristics of possible alpha decays of dysprosium
isotopes to the lowest excited levels of the daughter nuclei.}
\begin{center}
\begin{tabular}{|l|l|l|l|l|l|}
 \hline
  Alpha decay                       & Excited level         & Detection     & $\lim S$  & Experimental              & Theoretical    \\
  $Q_{\alpha}$ (keV) \cite{Aud03}   & of daughter           & efficiency    & ~         & limit on                  & estimation of  \\
  $\delta$ (\%)  \cite{Boh05}       & nuclei (keV),         & of the $\gamma$ & ~       & $T_{1/2}$ (yr)            & $T_{1/2}$ (yr) \\
   ~                                & yield of $\gamma$     & quanta        & ~         & ~                         &  \cite{Poe83} \\
   ~                                & per $\alpha$ decay    &  ~            & ~         & ~                         & ~ \\
  \hline
  $^{156}$Dy$\rightarrow$$^{152}$Gd & $2^+$, 344.3          & 2.1\%         & 31        & $\geq3.8\times10^{16}$    & $2.3\times10^{34}$ \\
  1759(6)                           & 0.962 \cite{Art96}    & ~             & ~         & ~                         & ~ \\
  0.056(3)                          & ~                     & ~             & ~         & ~                         & ~ \\
    \hline
  $^{158}$Dy$\rightarrow$$^{154}$Gd & $2^+$, 123.1          & 0.9\%         & 31        & $\geq1.3\times10^{16}$    & $1.4\times10^{68}$ \\
  879(3)                            & 0.455 \cite{Rei09}    & ~             & ~         & ~                         & ~ \\
  0.095(3)                          & ~                     & ~             & ~         & ~                         & ~ \\
    \hline
  $^{160}$Dy$\rightarrow$$^{156}$Gd & $2^+$, 89.0           & 0.4\%         & 230       & $\geq8.5\times10^{15}$    & $5.4\times10^{126}$ \\
  439.2(13)                         & 0.203 \cite{Rei03}    & ~             & ~         & ~                         & ~ \\
  2.329(18)                         & ~                     & ~             & ~         & ~                         & ~ \\
    \hline
  $^{161}$Dy$\rightarrow$$^{157}$Gd & $5/2^-$, 54.5         & 0.2\%         & 83        & $\geq3.5\times10^{16}$    & $1.3\times10^{147}$ \\
  344.6(12)                         & 0.075 \cite{Hel04a}   & ~             & ~         & ~                         & ~ \\
  18.889(42)                        & ~                     & ~             & ~         & ~                         & ~ \\
    \hline
  $^{162}$Dy$\rightarrow$$^{158}$Gd & $2^+$, 79.5           & 0.3\%         & 113       & $\geq1.0\times10^{17}$    & $1.7\times10^{1414}$ \\
  85.0(12)                          & 0.142 \cite{Hel04}   & ~             & ~         & ~                         &  ~ \\
  25.475(36)                        & ~                     & ~             & ~         & ~                         & ~ \\
    \hline

\end{tabular}
\end{center}
\label{tb5}
\end{table}
The theoretical
half-lives were found very big. Indeed, the predicted half-life is
$T_{1/2}=2.3\times10^{34}$ yr even for the decay with the biggest
energy release ($^{156}$Dy $\rightarrow~^{152}$Gd,$~2^{+},~$344.3
keV).

Theoretical estimates give no hope for realistic observation of
alpha decays of all naturally abundant dysprosium isotopes to the
excited level of daughter nuclei. At the same time the detection of the 
$^{156}$Dy decay to the ground state of $^{152}$Gd (supposing one
will apply crystal scintillators from dysprosium enriched in
$^{156}$Dy) looks possible. Indeed we have estimated the half-life
as $5\times10^{24}$ yr by using the approach \cite{Poe83}.
The important advantage of scintillation methods is an almost 100\%
detection efficiency to alpha particles and pulse-shape
discrimination ability \cite{Dan03,Bel07}. Development of
enrichment methods for dysprosium isotopes and radiopure crystal
scintillators containing Dy are requested to observe $\alpha$
activity of natural dysprosium.

\section{CONCLUSIONS}

The first experiment to search for $2\beta$ processes in
$^{156}$Dy and $^{158}$Dy was carried out in the underground Gran
Sasso National Laboratories of INFN by using ultra-low background
HP Ge $\gamma$ spectrometry. After 2512 h of data taking with a 322
g sample of highly pure dysprosium oxide (Dy$_2$O$_3$) limits on
double beta processes in $^{156}$Dy and $^{158}$Dy have been
established on the level of $10^{14}-10^{16}$ yr.

The search for the resonant $0\nu2\varepsilon$ capture in $^{156}$Dy
and $^{158}$Dy to excited levels of daughter nuclei was of
particular interest. Taking into account the strong dependence of
the process on the difference between the double beta decay energy
and the energy of an excited level, a more accurate determination
of the difference in atomic masses of $^{156}$Dy~--~$^{156}$Gd and
$^{158}$Dy~--~$^{158}$Gd isotopes is strongly required. A precise
study of the 1914.8, 1946.4, 1952.4, 1988.5 and 2003.8 keV levels
of $^{156}$Gd characteristics (spin, parity, decay scheme) is
important too.

Theoretical calculations of double beta decay of dysprosium are
scarce; in fact, the only known estimations for $^{156}$Dy (g.s.
to g.s. transitions) are
$T_{1/2}^{0\nu\varepsilon\beta^+}=7.0\times10^{27}$ yr (for the
effective neutrino mass of 1 eV) \cite{Rat09},
$T_{1/2}^{2\nu\varepsilon\beta^+}=(1.1-2.8)\times10^{26}$ yr,
$T_{1/2}^{2\nu2\varepsilon}=(1.8-4.5)\times10^{23}$ yr
\cite{Rat10}. The present study is the first
experimental attempt to search for double beta processes in
$^{156}$Dy and $^{158}$Dy, which both are rather interesting
isotopes for $2\beta$ experiments taking into account a possible
resonant double electron capture on excited levels of daughter
nuclei.

The sensitivity of the experiment can be advanced to the level of
 $T_{1/2}\sim 10^{23}-10^{24}$ yr by using enriched $^{156}$Dy and
$^{158}$Dy isotopes deeply purified from radioactive elements, and
increasing the exposure and detection efficiency by application of
multi-crystal HP Ge detectors.

The search for alpha decay of dysprosium isotopes to the excited
levels of daughter nuclei was realized as a by-product of the
experiment on the level of sensitivity $\lim T_{1/2}\sim
10^{15-17}$ yr. However, the obtained limits are very far from
the theoretical predictions even for the $\alpha$ transition of
$^{156}$Dy with the highest energy of decay ($T_{1/2}\sim 10^{34}$
yr). At the same time the detection of $\alpha$ decay of $^{156}$Dy to
the ground state of $^{152}$Gd (theoretical estimations of the
half-life are on the level of $\sim 10^{24}$ yr) looks reasonably
realistic by using crystal scintillators containing dysprosium
enriched in $^{156}$Dy. Such crystal scintillators could be also
used to search for double beta processes in $^{156}$Dy.

We have found that the Dy$_2$O$_3$ sample contains on the level of
several mBq/kg $^{176}$Lu, $^{235}$U and $^{226}$Ra; the
activities of $^{228}$Ra and $^{228}$Th are $\approx 0.18$ and
$\approx 0.16$ Bq/kg, respectively. The results of the $\gamma$
spectrometry are consistent with the data of the ICP-MS analysis for
radionuclides. The contamination of the dysprosium sample
can be explained by the production of lanthanides from monazite, an
ore with high concentration of uranium and thorium. Presence of
radioactive $^{176}$Lu can be due to the similar chemical properties
of lutetium and dysprosium, which provides some difficulties in
chemical separation of the elements.

\section{ACKNOWLEDGMENTS}

The authors from the Institute for Nuclear Research (Kyiv,
Ukraine) were supported in part by the Project
"Kosmomikrofizyka-2" (Astroparticle Physics) of the National
Academy of Sciences of Ukraine.

\end{document}